# A 149 minute periodicity underlies the X-ray flaring of Sgr A*

Elia Leibowitz

School of Physics & Astronomy and Wise Observatory

Sachler Faculty of Exact Sciences

Tel Aviv University

elia@wise.tau.ac.il

# Abstract

In a recent paper (Leibowitz 2017) I have shown that the 39 large X-ray flares of Sgr A* that were recorded by $Chandra$ observatory in the year 2012, are concentrated preferably around tick marks of an equi-distance grid on the time axis. The period of this grid as found in L1 is 0.1033 days. In this work I show that the effect can be found among all the large X-ray flares recoded by $Chandra$ and $XMM-Newton$ along 15 years. The midpoints of all the 71 large flares recorded between the years 2000 and 2014 are also tightly grouped around tick marks of a grid with this period, or more likely, 0.1032 day. This result is obtained with a confidence level of at least $3.27\sigma$ and very likely of $4.62\sigma$. I find also a possible hint that a similar grid is underlying IR flares of the object. I suggest that the pacemaker in the occurrences of the large X-ray flares of Sgr A* is a mass of the order of a low mass star or a small planet, in a slightly eccentric Keplerian orbit around the SMBH at the centre of the Galaxy. The radius of this orbit is about 6.6 Schwarzschild radii of the BH.

# 1. Introduction

In a recent article (Leibowitz 2017 - L1) it was shown that the 39 large X-ray flares of Sgr A* observed by the $Chandra$ space telescope in the year 2012, as presented in Table 1 of Neilsen et al (2013 - hereinafter N13), have an interesting statistical property. The set of the points on the time axis that mark the midpoints of each of these flares are tightly grouped around tick marks on the time axis that are separated from one another by integer numbers times a fixed time interval of length P=0.1033 days. Here we refer to the equi-distance between the grid points as the period of the grid.

Ponti et al (2015-hereinafter P15) have also analyzed the $Chandra$ data set of 2012, as well as the X-ray observations by this and the $XMM-Newton$ telescopes along the years 2000-2011 and the years 2013-2014. P15 used a different algorithm for defining distinguishable flaring events in the observed time series, termed blocks. There is a great deal of overlap between the 2012 39 flares of N13 and the 46 blocks in the same data set as defined by P15. The main difference between the 2 sets is that P15 considered as distinguishable blocks even those in groups of 2 or more outbursts of the source that occurred in immediate succession, with no time gap between the members of the group. In fact P15 themselves recognized each one of such small groups as an individual flare.

The analysis presented in L1 has not been applied on the 24 blocks recorded by $Chandra$ along the 12 years from 2000 to 2011 as presented in Table 4 of P15. It was thought that they are inadequate data basis for looking for the operation of a pacemaker with a period that is a fraction of a day. The period search that revealed the periodic grid underlying the N13 flare events was applied in L1 also on the 22 blocks of 2013-2014 presented in Table 6 of P15. No equi-distance grid on the time axis of any outstanding periodicity was revealed in that analysis of the timing of these blocks.

In this communication I want to show that reconsidering the full data set of all the large X-ray flares of Sgr A*, the operation of the pacemaker can be clearly identified throughout the entire 15 years time interval 2000-2014, with a statistical significance that is even much larger than the one found for the 2012 data set. In Section 6.1 it is also shown that in a very small published data set of IR observations one may find a hint that a similar pacemaker is involved also in the production of NIR flares of the object.

# 2. Data

The data used in our analysis here is the list of the 39 flares presented in Table 1 of N13 that were analyzed in L1. The parameters of an additional flare in the year 2012 were extracted from the upper left frame in Figure 1 in the paper by Zhang et al (2017). We also use the 24 blocks recorded by $Chandra$ between the years 2000 and 2011 as listed in Table 4 of P15. We further used the 22 blocks of 2013-2014 presented in Table 6 of P15, as well as the 23 blocks recorded by $XMM\ Newton$ between the years 2001-2014 presented in Table 7.

Following P15, we consider all subsets of 2 or more successive blocks that have no gaps between them as single flares, as mentioned above. We then omitted all the small flares with fluence smaller than $10 \times 10^{-10} erg\ cm^{-2}$. They include one of the 2012 flares. The data base of our analysis in this work is then 71 large X-ray flares recoded by $Chandra$ or $XMM-Newton$ in the 15 years 2000-2014. We denote these flares as Set C. It is here divided into 2 subsets: Set A is the 39 flares of 2012 and Set B consists of 32 flares, the 23 flares of 2000-2011 + the 9 flares of 2013-2014.

## 3. Method

The method to unveil a grid of points on the time axis of a significant period, of the nature described above, for a set of N flares, if one exists, is presented in L1. Here we describe it again, and add one more step in the data analysis procedure.

Both N13 and P15 give the beginning and the end times of each flare as determined by their algorithms. For each flare we define its time of occurrence as the midpoint between these two times. We thus obtain on the time axis a set of time coordinates t(i) of N "events". We have converted the MJT times given in N13 and P15 to HJD times. This conversion improved the statistical significance of the results that we obtained, as will be discussed in Section 5.1. The HJD times of each of these events relative to the time $t(1) = 2451844.6883$ are listed in Table 1 and are presented graphically in Figure 1. The vertical lines in the figure delimit the events of set A. The arrows point at gaps in the otherwise more dense distribution along the time axis of the 2012 data points. All 4 gaps are in the range of 40 to 62 days. See Section 4 for an explanation of their significance.

| | | | | | |
|---|---|---|---|---|---|
| 0 | 312.1713 | 574.6402 | 576.0045 | 576.4139 | 578.4679 |
| 579.0831 | 579.5993 | 706.2493 | 1348.9707 | 1736.9417 | 1737.7548 |
| 2089.0727 | 2350.0525 | 2350.4374 | 3124.9345 | 3125.2410 | 3125.7872 |
| 3125.8559 | 3484.9861 | 3805.2781 | 3806.5637 | 3810.1674 | 4122.2568 |
| 4122.4428 | 4162.3015 | 4162.3440 | 4162.4015 | 4162.4960 | 4204.3448 |
| 4204.5018 | 4209.9466 | 4214.5129 | 4214.8414 | 4215.1379 | 4215.9640 |
| 4223.6924 | 4282.8253 | 4283.0038 | 4284.3677 | 4286.0201 | 4286.7301 |
| 4287.3560 | 4288.2085 | 4289.3329 | 4289.8359 | 4295.2086 | 4296.8380 |
| 4299.1388 | 4300.1577 | 4302.9565 | 4362.9940 | 4364.0144 | 4370.8634 |
| 4372.0527 | 4372.9051 | 4373.659 | 4379.2325 | 4381.0549 | 4386.1330 |
| 4386.5370 | 4387.3869 | 4672.2321 | 4704.9290 | 4749.4965 | 5055.3327 |
| 5055.8167 | 5056.0107 | 5056.869 | 5085.1099 | 5106.3831 | |

Table 1: Times of midpoints of 71 large X-ray flares recoded by $Chandra$ and $XMM-Newton$ along the years $2000 - 2014$.  t=HJD-2451844.6883

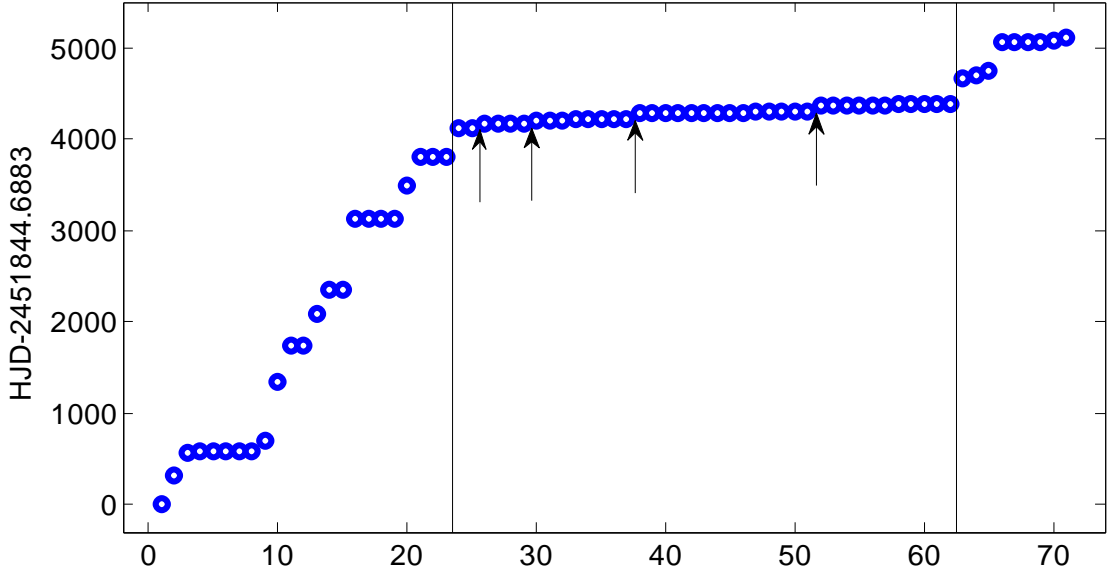

Figure 1: Times of midpoints of 71 large X-ray flares recorded by $Chandra$ and $XMM-Newton$ along the years 2000 – 2014. t=HJD-2451844.6883 (data of Table 1). Points between the two vertical lines are set A, the 2012 data. Arrows point at time intervals, all in the range of 40 to 62 days, between successive points. See text for further explanation.

We consider a set of $n_f$ frequencies $f(j)$ that are equally spaced in the frequency interval that corresponds to the period interval [0.05-0.5] day. In L1 the period interval that was considered was [0.02-0.5] day, but see also Section 4 and Figure 7a1 below. For each of the $q(j) = 1/f(j)$ within the search interval we consider a grid of points on the time axis, the distance of each one of them from the first event (or from a few different points around it - see below) in the time series considered is $r \times q(j)$, where $r$ is an integer.

We now calculate for each of the times $t(i)$ of the N events its distance from the nearest point of the $q(j)$ grid as a fraction of the value of the period q(j). In other words, we find the distance of t(i) from the nearest grid point that is equal or smaller than q(j)/2. We therefore define

(2) $$d[i, \mathrm{q}(j), t(0)] = dif\left[\frac{t(i)-t(0)}{\mathrm{q}(j)}\right]$$

Here $dif$ is the decimal fraction $d$ of the distance of t(i), expressed in units of q(j), from its nearest integer number : $-\frac{1}{2} \leq d \leq \frac{1}{2}$.

For each value of q(j) we compute the variance of the ensemble of the $d$ values corresponding to the given N t(i) times:

(3)

For each of the q(j) values, the value of $s^2[q(j)]$ is computed 2n times with 2n different reference times. They are t(1) itself and another n equally spaced points on the time axis that cover half a cycle of the period q(j) on each side of t(1). The grid with respect to the $2n + 1$ point, which is $t(1) + q(j)/2$, is the same as the grid measured from $t(1) - q(j)/2$, only with different r values. Therefore it does not have to be considered. In our calculation we took $n = 2$. Taking $n \geq 3$ yields very similar results.

The variance of the observations with respect to the tested period q(j), $S^2[q(j)]$, is the smallest value of $s^2[q(j)]$ among those obtained for all 2n zero times considered. In a plot of the standard deviation S vs. $f = \frac{1}{q}$ each S value expresses the dispersion, in phase units, of the times of the N events around tick marks of a grid on the time axis of the corresponding $q = \frac{1}{f}$ periodicity. We refer to this plot as the Frequency-Dispersion Diagram (FDD). The value q(j) that minimizes $S^2[q(j)]$ is the period P of the grid that with respect to its tick marks, the observed events are grouped together most tightly. The corresponding beginning time is taken as an initial, reference time, associated with this period.

As seen in Figure 1, there are very large gaps between the recorded events, especially between those along the years 2000-2011 and 2013-2014 which are up to 4 orders of magnitudes larger than the periods that are of our concern here. Therefore, the integer numbers r mentioned above, i.e. the number of cycles of the periods within these gaps, is very large. Even very small difference between 2 tested frequencies may introduce a considerable change in the corresponding r values. This is reflected in the structure of the FDDs as large fluctuations, as shown in Figures 2a and 2b. These are FDD plots in the frequency band [9.66-9.7] $day^{-1}$, corresponding to the period interval [0.1031-0.1035] day, calculated for the 71 flares of set C. Frame a is with 500 frequencies and frame b with 1500 frequencies within this interval. It is apparent in the 2 figures that the period of the lowest minimum in the FDD, as well as the corresponding S value, is quite sensitive to the particular number $n_f$ of tested frequencies within the frequency search interval. In order to overcome this discontinuity in the structure of the FDD we apply the running mean operation on the FDD function and adopt the period corresponding to the minimum of this curve as the best outcome of the analysis.

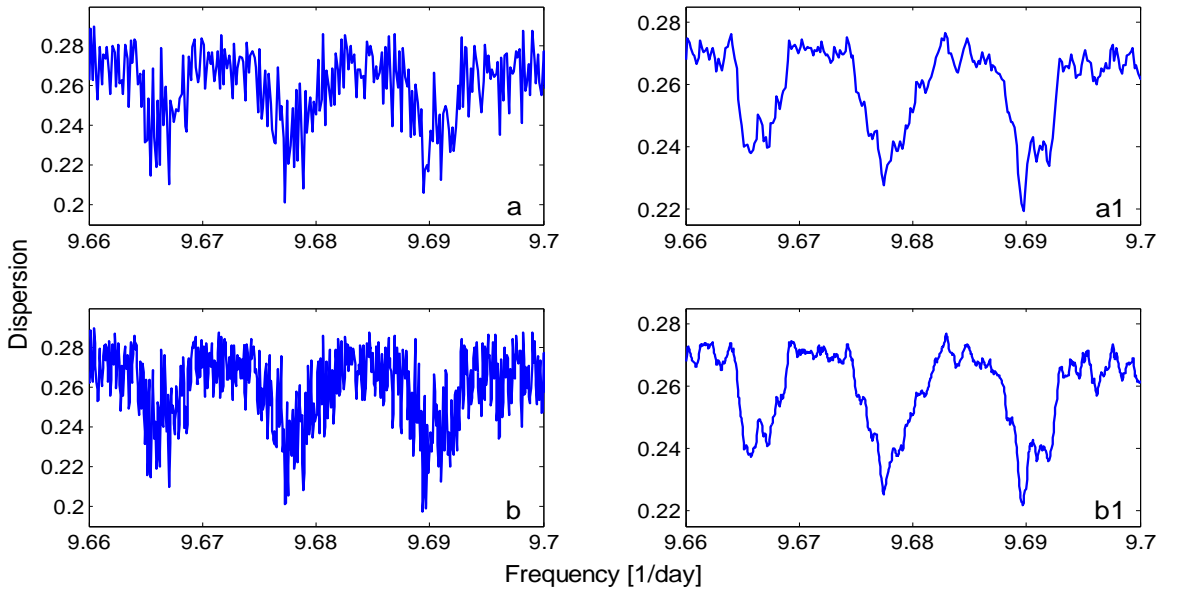

Figure 2: Frequency-Dispersion Diagrams for the midpoints of the 71 flares recorded in the years 2000-2014. (a) Number of sampled frequencies in the displayed band $n_f = 500$. (b) $n_f = 1500$ (a1),(b1) Running mean of the corresponding FDDs in frames a and b with running window 0.00072 $day^{-1}$.

Figure 2a1 presents the running mean of frame a with a running window width of 0.00072 day$^{-1}$. Figure 2b1 is the running mean of frame b with the same running mean widow. Figure 2 demonstrates that for large enough $n_f$, the period of the minimum point found in the running mean of the FDD is nearly independent of the number $n_f$. From here on we refer to the running mean function as the FDD of the time series, unless stated otherwise.

## 4. Results

Figure 3a is the FDD computed for the 39 flares of set A, those recorded in the year 2012, in the frequency range [2-20] $day^{-1}$, corresponding to the period range [0.05-0.5] day. Here the number of frequencies considered within this range is 100000 and the running mean operator was applied with a window width of 0.00198 $day^{-1}$. Frame b is the same for the 32 flares of set B. Here the widow width is 0.0126 $day^{-1}$. Frame c is the same for the 71 flares of the combined set C with window = 0.00072 $day^{-1}$. Frame a1, b1 and c1 are zooms on a narrow band of the frequency range, containing the outstanding minimum features in the corresponding frames to the left. The vertical lines mark the frequency F1=9.6898 $day^{-1}$, corresponding to the period P1=0.1032 day. We note that extending the search interval in frames a and c all the way down to $p = 0.01\ day$ ($f = 100\ day^{-1}$) does not reveal in the FDD any additional minimum points that are equal or deeper than the ones seen in the displayed diagrams. For the range [2-50] $day^{-1}$ this can be seen in Figure 7a1.

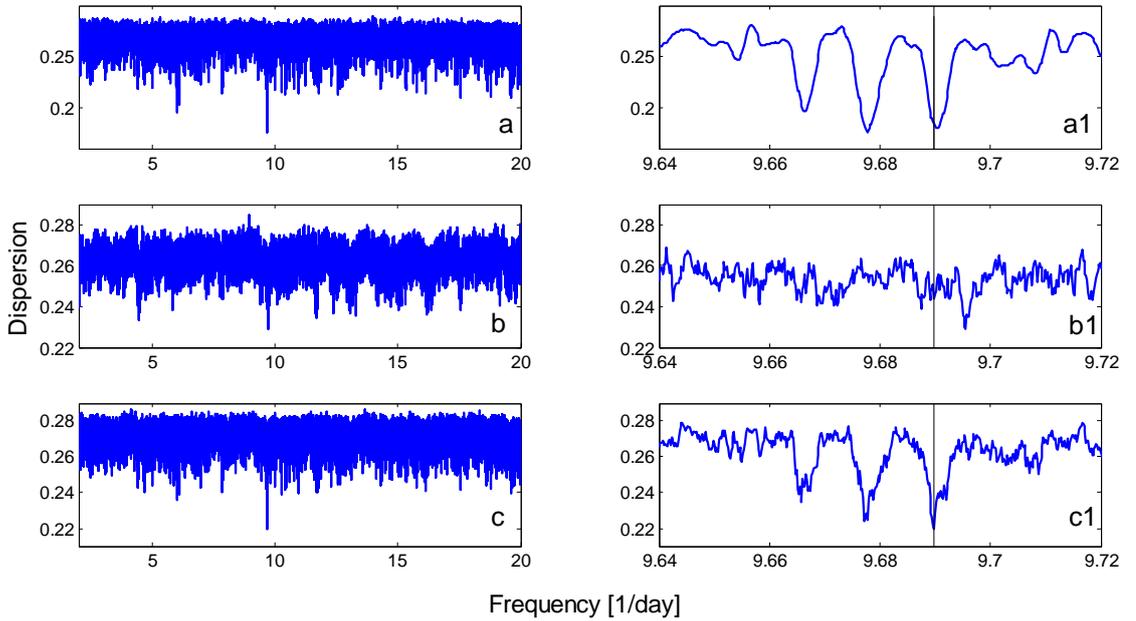

Figure 3: (a) Frequency Dispersion Diagram for the 39 flares of set A in the frequency range [2-20] $day^{-1}$ with $n_f = 100000$ and running window 0.00198 $day^{-1}$. (b) Same for the 32 flares of set B with running mean widow 0.00126 $day^{-1}$. (c) Same for the 71 flares of the combined set C with running window $0.00072 day^{-1}$. (a1), (b1), (c1) Zooms on a narrow frequency band around the deepest minima in the corresponding frames to the left. Vertical lines mark the frequency F1=9.6898 $day^{-1}$ corresponding to the period P1=0.10320 day.

The frequencies of the minimum points of the 3 components of the triple feature seen in frames a1 and c1 are F1=9.6898, F2=9.6775, F3=9.6655 $day^{-1}$, corresponding to P1=0.10320, P2=0.10333, P3=0.10345 day, are aliases of one another. They are due to the 4 relatively large gaps between 5 subgroups of the much denser points of set A. These gaps are indicated by the 4 arrows in Figure 1. As pointed out in Section 2, all 4 gaps have a similar size of about $50\ days$. The period P1 in the triple feature in the FDDs is different from P2 due to the fact that the number of cycles of P1 in each gap, where no events have been recorded, is larger by 1/2 than the number of cycles of P2 in this gap. The third component of the triple feature is understood in a similar way. Note that there is no similar triplet feature in the FDD of set B that, indeed, does not contain the origin of these aliases in sets A and C. The minimum point in the FDD of set B is at the frequency F4=9.6955 $day^{-1}$, corresponding to the period P4=0.10314 day.

Applying the period search operation on a very high density of sampled frequencies we find that the smallest dispersion S=0.18831 is obtained at P1=0.103203 day=148.612 min. However, one cannot rule out the possibility that the true period is P2=0.103334, the one corresponding to the central minimum of the triplet feature.

The distance between the two right minima in the triplet feature is $\Delta f \cong 0.02\ day^{-1}$. The corresponding $\Delta P \cong 1.065 \times 10^{-4}\ day \cong 9\ sec$. The uncertainty in the reference time t(0) is estimated by one quarter of the value of the period, namely by $0.025\ d \cong 36\ min$.

Assuming that P1 is the period, the ephemeris of the pacemaker ticks on the time axis is then:

$$HJD = 2451844.7141 \pm 0.025 + (0.103203 \pm 0.0001)E \quad E = 0,1,2,....$$

It should perhaps be emphasized here again, that although we are using the word "period" and presenting this ephemeris, we are not referring to any periodicity in the timing of any measureable phenomenon. In fact, the occurrence of midpoints of large X-ray flares of Sgr A* is clearly not periodic, certainly not with the period discussed in this paper. This is evident by the fact that most of the points on the "periodic" grid on the time axis are not marking times of mid-flares, most of them do not host any flare event even in their close neighbourhood. Accordingly, this ephemeris does not enable one to predict an occurrence time of any flare. It has only a statistical meaning. It provides, within the above stated uncertainty, a series of HJD dates such that the midpoints of future large X-ray flares of Sgr A* will be distributed around a few of them as centres, with standard deviation of about 28 min. At the present level of our knowledge of the Sgr A* system, the identity of these few dates is unpredictable. Confirmation of this ephemeris can therefore be made only statistically, namely, on the basis of a certain number of large flares that will be recorded in the future. See also section 6.2 for further discussion of this point.

Most recently Mossoux and Grosso (2017 – hereinafter MG) have published a new compilation of 99 X-ray flares of Sgr A* recorded between the years 1999-2015 by the $Chandra$ and $XMM-Newton$ telescopes. We have not used the data of the 8 $Swift$ flares presented in this paper because of the low resolution in their timing. MG executed their own selection of flares from the available time series recorded by the two X-ray telescopes, as well as their own determination of the beginning and end times of the flares. So while their data base is obviously largely identical to that of N13 or P15, the data reduction process of MG is independent of the two earlier compilations. The MG list includes 7 flares of which either the beginning or the end times are not known because they were out of the time interval covered by the observations. In the MG paper there is also no classification of the events into small or large flare types. We applied our search routine on the entire ensemble of the 99 flares, with their associated time parameters just as presented in Tables A.1 and A.2 in MG paper. We found that the second deepest minimum in the resulting periodogram is at the period p=0.10331.

## 5. Significance

### 5.1 Qualitative

In order to evaluate the significance of our findings in the timing of the X-ray flares of Sgr A* we need to estimate their False Positive Probability (FPP). This is the probability that the period P1 found for the grid underlying the data is due to randomness in the distribution of times of the midpoints of the flares, rather than to a genuine regularity associated with the radiation source.

On a qualitative level we note the nearly perfect coincidence between the P value of the 3 lowest minimum points found among the 71 flares recorded in the 15 years between 2000 and 2014, and the 39 flares recorded in the year 2012, as seen in Figure 3. These data sets are of course not independent of each other as the latter one is a subset of the former. Nevertheless, the fact that adding the 32 independent times of 2000-2011 and of 2013-2014 to the 39 times of 2012 leaves the P value unchanged to within 0.0001 day≅11 sec, which is 1/1000 of the period, may add some credence to the reality of the P value. A quantitative related evidence is presented in Section 5.2.1.2.

As mentioned above, the numbers in Table 1 that we are analyzing here are HJD dates. The dispersion of the 71 points of set C around the tick marks of the P1 grid is S=0.21965. We show in Section 5.2.1.1 that this result is obtained at a $3.27\sigma$ level of confidence. If we consider the MJD of the same 71 events, as they are given in N13 and P15, rather than their HJD, the most tightly grouping of them is still around tick marks of the P1 grid but the dispersion is now S=0.22130. This reduces the confidence level of the found periodicity to a mere $3.04\sigma$ value.

The improvement in the statistical significance of the finding of the P1 periodicity achieved by the MJD to HJD conversion is a qualitative indication that the periodic grid on the time axis such that some of its tick marks are centres around which the arrival times of the flare midpoints are tightly grouped together, may indeed be related to some external reality and not to a merely random coincidence.

### 5.2 Quantitative

#### 5.2.1 Simulations Type I

##### 5.2.1.1 Set C

On a quantitative level, we estimate the FPP of our finding by way of simulations, as was done in L1. A simulated set of N pseudo-observed (PO) events is created as follows: We consider the N days of the real observed flares. For each one of them we replace the fraction of the day at which the midpoint of the flare occurred, expressed in decimal units, by a random number selected from a rectangular distribution over the [0,1] interval. The random number generator that we used is that of the MATLAB computing environment. A simulated set so created preserves all the temporal content of the real data except the hour in the day of the midpoint of the real flares. We apply on this set of N PO events the same search routine that was applied on the real data and find the q value of the grid, with respect to which the dispersion s takes its minimum value. This is performed on a set of K samples of N PO events. The fraction k/K, where k is the number of PO sets for which $s \leq S$ serves as an estimator of the desired probability.

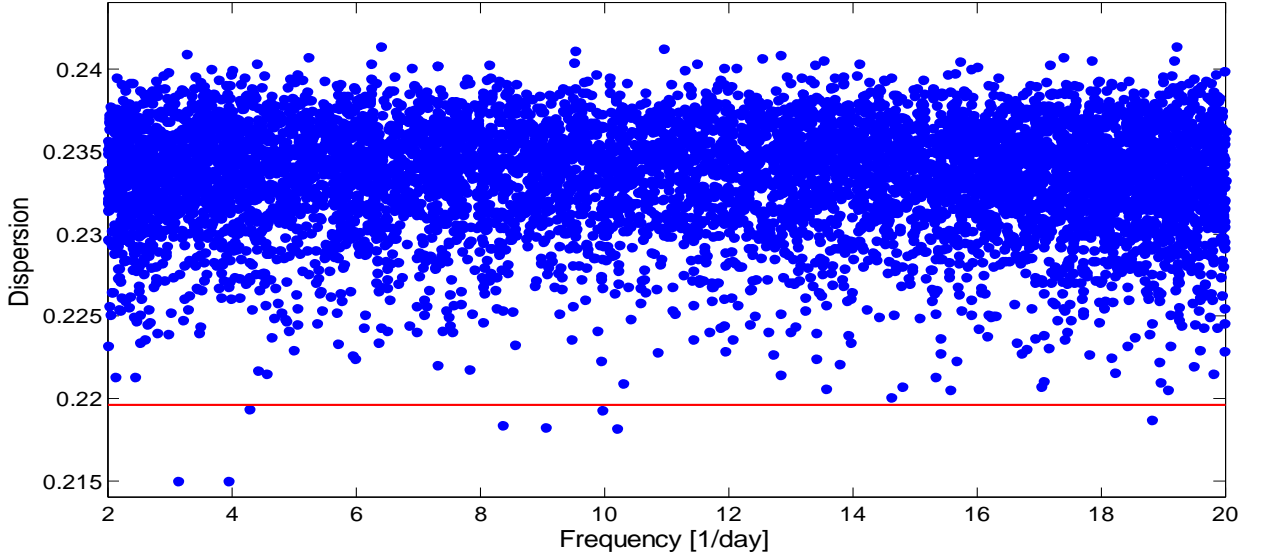

Figure 4: Results of simulations. The minimum s values vs. the frequencies f at which they are found in a sample of 7500 sets of 71 pseudo-observed events, simulating the observed data set C. Horizontal line marks the dispersion 0.21965 of the midpoints of the observed 71 flares around tick marks of a grid on the time axis of the period P1=0.1032 day.

Figure 4 is a plot of s, the minimum dispersion values found in 7500 sets of 71 PO events, simulating the real data of set C, vs. f, the frequencies at which the corresponding s values were found. The horizontal line marks the dispersion 0.21965 of the midpoints of the 71 observed flares of set C around the best fit grid of the period P1=0.1032 day. There are 8 points of the simulated data that fall below the horizontal line. We therefore estimate the False Positive Probability of finding as a random event a grid of the P1 period with respect to which the times of set C has the dispersion S as $FPP_1 = \frac{8}{7500} = 0.0010667$. The corresponding significance level of our finding in the 71 large flares recorded along the years 2000-2014 is $L_1 = 3.27\sigma$.

**5.2.1.2. Sets A and B**

We have created 3000 sets of 39 PO events, simulating the data of set A, as done for set C. We found 36 sets with an s value that is smaller than S=0.1765 of the real data. The FPP of the finding of the P2 periodicity in the best fitted grid to the set A is therefore estimated as FPP(A)=0.012. In a similar way we find in a sample of 200 sets of 32 PO events simulating the set B data 32 cases with s smaller than S=0.2290 that is found for the real data. From this we estimate FPP(B)=0.16

The data set B is a series of times that are independent of the set of times A, except perhaps for the fact that the same two telescopes were used in the collection of the data of the 2 sets. The width of our frequency search interval is 18 $day^{-1}$. The probability that the frequency F4=9.6955 of the grid best fitted to the events of set B will be found at random as close to F2=9.6775 of the grid best fitted to the independent set A is FPP(AB) = 2(F4-F2)/18=0.002. An estimate of the FPP of F4 to be a random result of our search routine is the product of the 3 numbers FPP(A), FPP(B) and FPP(AB), namely, $FPP_2 = 3.84 \times 10^{-6}$.

The result F4 found for set B is consistent with identifying the frequency associated with this set with the frequency F1 of set C or with F2, the frequency of set A . We show this as follows. From the 71 tick marks defined by the ephemeris given above we consider the 32 marks that correspond to set B. To each one on these times we add a number, selected randomly (with repetitions) from the ensemble of the 32 differences between the set B times and the corresponding tick marks of the F4 grid found by our search routine for this set. Repeating this procedure 100 times, each time with a new random selection, we create 100 sets of 32 events that simulate set B. By their very construction, all these sets are built upon the F1 grid. Applying our search routine on these PO set B series we find 50 best fit grids with periods that are as far or farther from F1 as F4 is.

The $FPP_2$ number can be translated to $L_2 = 4.62\sigma$ level of confidence in the results of this work. An explanation of the large difference between $L_1$ and $L_2$ is given in Section 7.1.

### 5.2.2 Simulations Type II

We performed also a second type of simulations of the observed data set C as follows: From Table 3 of P15 we take as 1 per day the mean flaring rate of Sgr A* over the 15 years of our interest. Li, Yuan and Wang (2017) quote the number 2 per day for the mean X-ray flares frequency but it seems that for large flares, the number 1 per day is more consistent with the number of large flares that have been actually observed over the 15 years 2000-2014. Also P15 suggest that after the year 2014 there might be an increase in that rate. However, these two variations on the flare frequency that we have adopted do not affect significantly the simulated process that we are here describing. The observations discussed in this paper extend over the period of some 5510 days. We therefore select 5510 random numbers from a rectangular distribution over the time interval of these 5510 days. We now find which of these numbers fall within the 149 observational windows of the X-ray satellites as presented in Tables 4-7 in P15. We consider these numbers as midpoints of PO flares. As we have done for the type I simulations, we perform on these type II simulated sets of PO events the search routine performed on the real data. Here, in the running mean step of the process, we used a widow width of 0.00072 day$^{-1}$.

Figure 5 is a plot of the minimum dispersion values s found in a sample of 3000 sets of Type II PO events, vs. the number of events in each set so constructed. The horizontal line marks the S value of the observed set. There is a clear correlation between the s value of the PO sets and the number n of events in each one of them, with a correlation coefficient 0.63493 . The linear regression line between these two parameters is also presented in the figure.

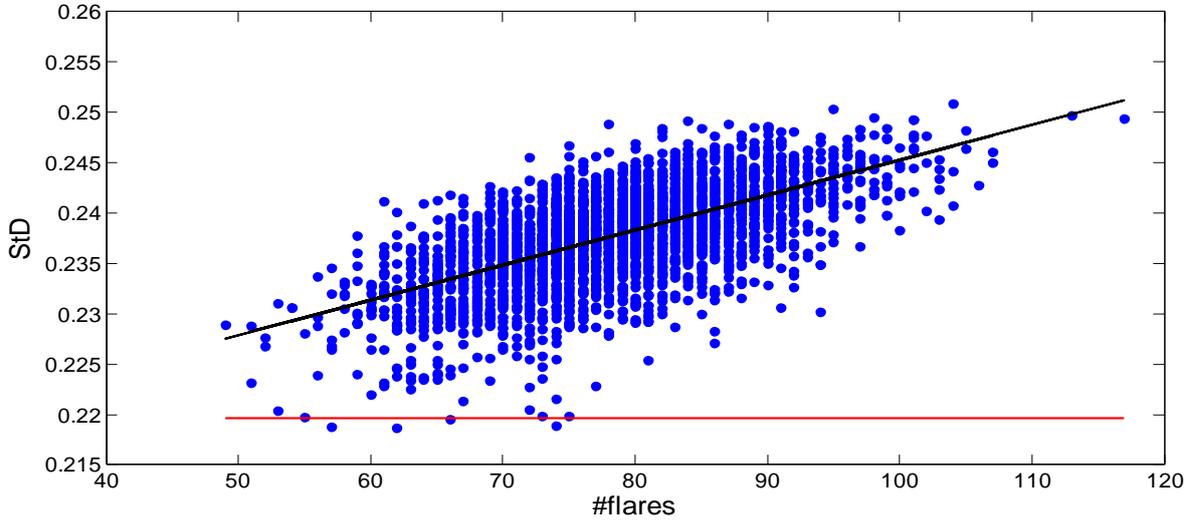

Figure 5: Minimum dispersion values s vs. the number of PO flares in each one of a sample of 3000 type II simulations of the midpoints of the observed large flares of Sgr A*, along with the regression line between these two parameters. Horizontal line marks the minimum dispersion value S found in the timing of the 71 observed flares.

There are 4 points below the horizontal line in the figure, namely, 4 sets for which the s value is found to be smaller than the S value of the observed data. The probability 4/3000 makes the S number statistically significant at a $3.21\sigma$ confidence level. If we remove from the simulated data the systematic trend expressed by the regression line, there remain only 3 cases with dispersion smaller than that of the observed data. This corresponds to a $3.29\sigma$ level of confidence.

As we shall discuss in the next section, there is no dependent variable in the time series that we are analyzing in this work. Therefore, the rejection of the null hypothesis allows us to conclude this section with the following statement:

The midpoints of the 71 large flares recorded by $Chandra$ and $XMM$ in the 15 years 2000-2014 are grouped preferably around tick marks of an equi-distance grid of points on the time axis that has the periodicity $P1 = 0.103203\ day$. The standard deviation of the distribution of the midpoints around the tick marks is 0.18831 of the period, or about 28 min. The reality of this result can be accepted at a $L_1 = 3.2\sigma$ level of confidence or higher, and is also supported by some qualitative considerations. As we shall argue in Section 7.1 the value $L_2 = 4.62\sigma$ is a realistic estimate of the level of confidence in our results.

## 6. The IR Connection

### 6.1 Analysis of 5 NIR flares

A strong correlation has been found between the variability of the IR radiation of Sgr A* and that of its X-ray radiation (Li, Yuan, Wang, 2017 and references therein) . It has been even suggested that each X-ray flare is accompanied by a NIR flare (Meyer et al, 2009, hereinafter M2009).

An example of the IR variability can be found in the set of 5 NIR flares in 2004, presented and newly analyzed by M2009. We have applied our period search technique on the midpoints of these 5 NIR flares, extracted from Figure 1 of M2009. The FDD was computed over the range [0.01-1] day. It is dominated by 7 outstanding minima, all confined within the period interval [0.067-0.105] day. They appear in the [0.01-1] period range together with their 7 higher harmonics.

The first 4 IR flares fell within one of the observational window of *Chandra*. In the P15 list of large X-ray flares there are no counterparts to the first 3 of them. The 4th IR flare coincides, within less than 9 minutes, with X-ray flare No. 10 in Table 1 of this paper. The 5th flare is outside the observational windows of the two X-ray telescopes. Considering all this and the fact that the observations analyzed anew by M2009 extend over merely 2 days, the rather feeble results of the above analysis may still be regarded as some additional, although quite weak qualitative evidence for the reality of the grid of the P periodicity found in the X-ray data, that comes from an entirely different mode of observations. The results of the analysis of the very small set of IR data may also be appreciated as hinting about the possible involvement of a pacemaker with the generation process of the IR flares of the system that is similar to the one found in the X-ray data. Naturally, a real examination of the possibility that the IR flares of Sgr A* are also modulated by a periodic pacemaker requires a much richer data set of IR observations.

## 6.2 NIR and X-ray Light Curves

M2009 found that the power spectrum density (PSD) of time series of IR measurements may be represented by a power law of two different slopes. The break frequency that parts the two slopes in the power law representation corresponds to the period $p_{br} = 0.107\ day$. The analysis of M2009 was performed on a Structure Function, as defined in their paper, rather than on the PSD of the time series. It was however noted by these authors, that a break in the slope in the power law representing one of these functions translates to a break in the power law representing the other. The break period reported by M2009 is not too different prom the period 0.1032 d of the pacemaker that modulates the timing of the X-ray flares, as reported about in this work.

This apparent similarity between these two numbers may, however, be misleading. The time series analyzed in M2009 is fundamentally different from the subject matter of our analysis here. The M2009 result points at a characteristic of the IR variability that is different from the meaning of our result obtained for the X-ray flares. The time series analyzed by M2009 is a light curve. It consists of series of pairs of numbers, an independent variable, the time of each observation, and a dependent one, the IR intensity recorded at that time. Here we are considering a time series that contains only values of a single independent variable, the times of the midpoints of X-ray flares. This time series constitutes just a set of natural numbers. Therefore, while there is a lot of sense in discussing possible correlations in the dependent variable of the IR LC, as may possibly be revealed by the structure of the PSD, there is hardly any correlation that one may attribute to the series of numbers that are of our concern, other than their statistical properties such as the one revealed in this work.

This fundamental difference clarifies also the inadequacy of the PS or similar techniques for looking for the pattern of our interest in the X-ray time series. As was stressed in Section 4, here we are not claiming the discovery of any periodicity in the LC of the object. We therefore believe that the close similarity between the IR number 0.107 of M2009 d and our number 0.103 d is mostly a coincidence.

In order to further demonstrate this point we have constructed a LC of pairs of an independent and a dependent parameters of the X-ray data. According to P15, along the 15 years 2000-2014 there are 149 time intervals during which either the *Chandra* or the *XMM* observatories, or both, were monitoring the object. The time and duration of each of these 149 observational windows are given in Tables 4-7 in P15. In our LC construction, each one of these windows was covered by a grid of points with equal spacing of 0.01 d. Each point was assign the value 1 for the corresponding X-ray intensity. In all the 71 time intervals of recorded large X-ray flares within these windows, the 1 values were superposed by numbers with equal spacing of 0.002 d, that establish an isosceles triangle over the flare duration time, whose area is proportional to the fluence of the corresponding flare. Figure 6 presents the LC so obtained, and zooms on 6 subsections of it.

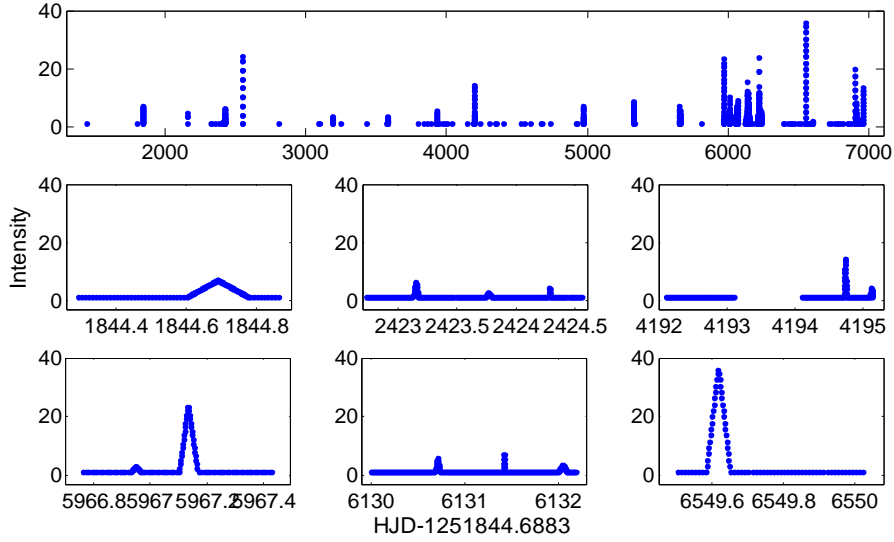

Figure 6: Synthetic light curve constructed around the 71 large X-ray flares of Sgr A* recorded between the years 2000-2014, and zooms on 6 small subsections of it. See text for explanations.

Figure 7a1 presents again the FDD of the 71 real flares computed in the frequency range [2-50] $day^{-1}$, corresponding to the period interval [0.02-0.5] d, on 300000 sampled frequencies and running mean window=0.00072 $day^{-1}$. Frame a2 is the power spectrum of the LC presented in Figure 6 in the frequency range and sampling rate as in frame a1, binned into 30000 bins. Notice that the highest peak in frame a2 is not at the frequency of the minimum point in frame a1.

Following M2009, we find by Least Squares the break frequency $f_{br}$ and the two slopes $\gamma$ and $\beta$ of a 2 slopes power law representing the systematic trend apparent in frame a2. The solid line in Figure a2 is the power law representation with the parameter values $f_{br} = 10.05\ day^{-1}$, ($p_{br} = 0.0995 day$), $\gamma = 0.375$, $\beta = 2.3$. We found that no power law with one single slope may reasonably be considered a fair representation of the PS. We note also that very similar results are obtained if we consider a LC consisting of just the triangles mentioned above that simulate only the 71 flares.

We have constructed a second synthetic X-ray LC around one of the simulated sets of our type I simulation, of which the best fitted grid cycle is p=0.3184 day (f=3.1411 $day^{-1}$), with s=0.2159 < S=0.21965. The FDD for this set of 71 simulated events is displayed in Figure 7b1, and the PS of this LC is shown in frame 7b2, as in the a frames. Here again the frequency of the highest peak in frame b2 is not the frequency of the minimum point in frame b1. The solid line is the best fitted double slope power law presentation, with the parameter values $f_{br} = 10.25\ day^{-1}, (p_{br} = 0.0976\ day), \gamma = 0.375, \beta = 2.35$.

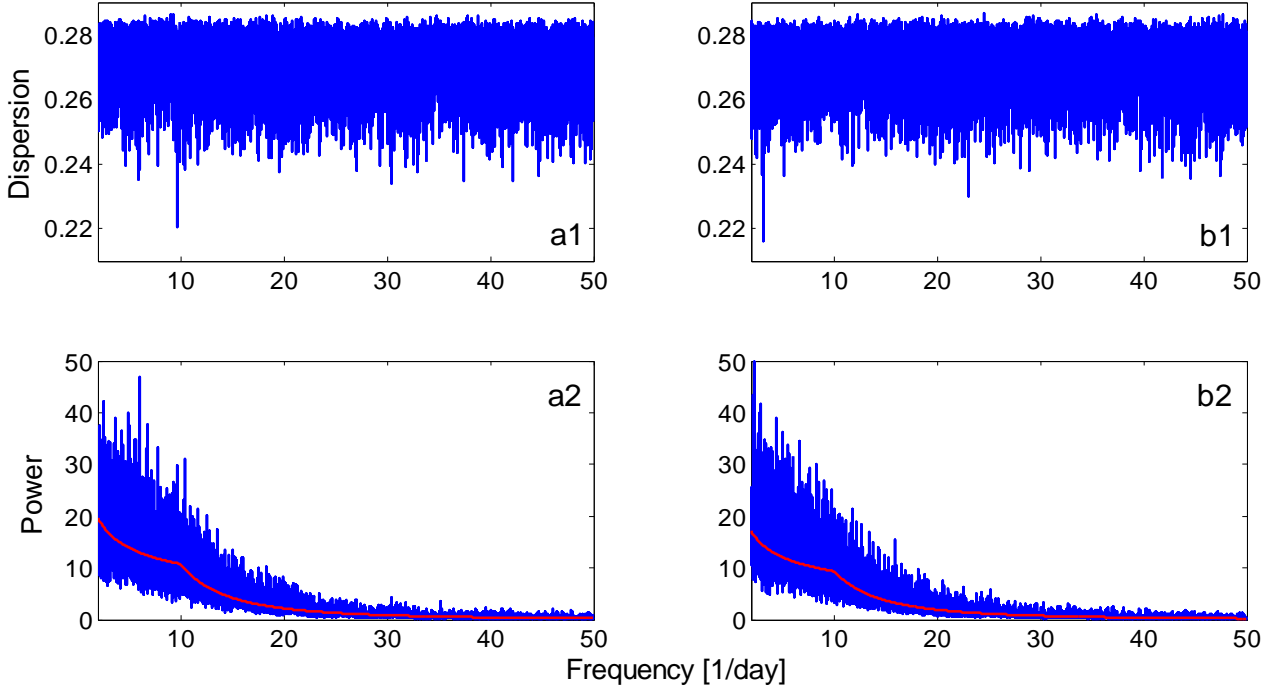

Figure 7: (a1) Frequency Dispersion Diagram of the 71 large X-ray flares computed over the frequency range [2-50] $day^{-1}$, corresponding to the period range [0.02-0.5] day, on 300000 frequencies, on which the running mean operator was applied with window=0.0072 $day^{-1}$. (a2) Power spectrum of the light curve presented in Figure 6, on the same frequency interval and the same sampling rate as those of frame a1, binned into 30000 bins. Solid line is the best fit double slopes power law representing the systematic trend of the PS. Frames b are similar to frames a, for one of the 8 sets of PO flares for which a grid of f=3.1411 $day^{-1}$, or p=0.3184 day, was found such that s=0.2159, the dispersion of the PO points around its tick marks, is smaller than S=0.21965, the dispersion of the observed points around tick marks of the P1= 0.1032 day grid.

We also performed similar exercises with LCs constructed around products of the type II simulations that generate sets of different numbers of PO events, as described in Section 5.3. In these synthetic LCs we found break frequencies in the range 9-11 $day^{-1}$ (periods 0.09-1.11 days) even for sets of PO midpoints for which the cycle length of the best fitted periodic grid were far away from this frequency interval.

Comparison of frames 1 and 2 of Figure 7 demonstrates vividly the difference between the nature of the phenomenon that we are investigating here and the more common analyses of timely behaviour of light curves of astronomical objects. While the signals of the 0.1032 day and the 0.3184 day periodicities are clearly pronounced in frames a1 and b1, they are hardly or not at all noticeable in frame a2 and b2. This is rightly so since Figure 6 is indeed not a LC of a periodic variable, nor is the LC constructed around the simulated set. The comparison between the two a frames or between the two b frames show the inadequacy of the PS and similar techniques for analyzing and exploring time series of the kind described in this work.

The outstanding minimum in Figures 7a1 is at the frequency revealed in this work of f=9.6898 $day^{-1}$ or p=0.1032 day, while in frame 7b1 the frequency of the minimum point is at f=3.1411 $day^{-1}, p = 0.3184\ day$. In spite of the large difference between the two grid periods, the PS of the LCs, as well as the break frequencies and the values of the 2 parameters $\gamma$ and $\beta$ of the power laws representing them, are about the same. This would indicate that the trend of the synthetic LC that we have constructed around the times of the observed flares, using their widths and fluences are determined mostly by the rate of large flares on time scale of a day and longer, and\or by the width and structure of the pulses and by the window function of the observations. The period of the pacemaker does not seem to play an important role in determining the structure of the PSD function.

I suspect that the same might be true to a large extent also for the PS of the IR light curve of the object. However, an attempt to examine this point and to discuss it any further is beyond the scope of this work.

## 7. Discussion

### 7.1 Statistical Confidence

Our finding that the set of 71 large X-ray flares of Sgr A* recorded along the years 2000-2014 are hiding the operation of a P1=0.1032 day pacemaker seems to be rather well established at the significant level of $L_2 = 4.62\sigma$, as shown in Section 5.2.1.2. This high significance level is not reflected in the estimate $L_1 \cong 3.27\sigma$, based on the simulations of all 71 flares of set C together, as presented in Section 5.1.2.1. The reason for this large difference is as follows.

The estimate of the FPP from set C is based on the dispersion of the real set, as compared to the dispersion in the PO sets. The dispersion in the real set is quite large (see below) therefore with the rather small number of merely 71 events, series of 71 random numbers, in the sense of our simulations, can accommodate with non negligible probability periodic grids with similar or smaller dispersions.

There is, however, a basic difference between the defining procedure of set C of the real data and the process involved in the finding of the 8 simulated C' sets that harbor a pacemaker grid with FPP values that are smaller than that of the grid associated with set C, as described in Section 5.2.1.1. The C' sets are those that contain as members 71 numbers with respect to which there is an equi-distance grid of some period P in the range [0.05-0.5] day that is found with small FPP. For all sets C' selected in this way the corresponding subsets A' and B' will also be best fitted with high probability to grids of similar P values. In contrast, membership in sets A and B of the real data, and hence also the membership in set C itself, is defined by the times of the detection of the large X-ray flares by the two X-ray telescopes $Chandra$ and $XMM-Newton$. In particular, in the definition of these observed sets, no reference is being made to any statistical property that the set of numbers so selected are or will be possessing. There is no a-priori reason why the best fitted grid to the 39 numbers of set A will have similar period to the period of the grid best fitted to the 32 dates of set B. For sets A' and B' such similarity is a direct consequence of the selection process of the sets as subsets of set C'. Therefore in evaluating the FPP of the results obtained from the real data, one has to take into account the implied stochastic element, under the null hypothesis, in the near equality P(A)=~P(B), as done in Section 5.2.1.2. This random element is missing from sets C' that simulate the set C of the real events as a whole. Hence the relatively low significance level of the results as estimated in Section 5.2.1.1.

The relatively high dispersion 0.21965 of the observed events around the tick marks of the pacemaker could be due to one or more possible reasons:

1. The result of our analysis is a statistical fluke after all.

2. The large flares of the system among which the operation of the pacemaker could be identified are recorded over a background of noisy emission that includes smaller flares in a decreasing ladder of intensity, either in their peak or fluence values. The pacemaker, if indeed exists in the system, is not regulating the production of all flares, as the ensemble of all flares does not show the regularity identified in the bright or very bright ones. The signal of the pacemaker in the distribution of flares along the time axis could perhaps become clearer if an objective criterion could be found that makes a more clear cut distinction between bright and faint flares.

3. We considered the midpoint of each flare as the point representing the occurrence time of the flare. Ponti et al (2017) have divided the time duration of 3 very bright flares into subsections, one of them includes the peak of the flare. In 2 of the three, VB1 and VB2 , the midpoint of the "Peak" subsection coincides with the midpoint of the flare as defined by the flare start and end times . In VB3 the midpoint of the "peak" section is about 400 seconds later than the flare midpoint. Also for some of the flares in set C we do not know the beginning or the end times of the flare itself since either one of these times occurred outside the observational window. For such flares the midpoint that we have been using is clearly not the real midpoint of the flare. It may well be that considering as the "events" to be analyzed the times of the peaks of the flares, or perhaps the times of some other identifiable feature in the flare morphology that is common to all of them, the dispersion of these events around the tic marks of the $P1=0.1032$ grid will become smaller.

4. The relatively large dispersion of the flare mid points around the grid tick marks could be inherent to the phenomenon itself. The centres around which the large flares of the system are grouped together may themselves be distributed along the time axis with periodic but non coherent regularity. This may be due to the effect of some of the processes that are involved with the generation of the X-ray flares that are not modulated by the P1 pacemaker.

## 7.2 Suggested interpretation

Assuming that a pacemaker with a periodicity of about 150 minutes has indeed operated in the Sgr A* system in the years 2000-2014 and possibly at the present time as well, we suggested in L1 an interpretation according to which the periodicity $P1=0.1032$ is the period of a material object in a nearly circular Keplerian orbit around the central blackhole. According to Abramowicz and Fragile (2013) , stable Keplerian orbits do indeed exist around non rotating blackholes, of radii that are larger than the radius of ISCO, the Innermost Stable Circular Orbit, which in the Schwarzschild metric is equal to 3 Schwarzschild radii.

The interaction of a SMBH with a stellar object at close proximity have been discussed in the literature in the last few years (e.g. Freitag, 2003; Miller et al, 2005; Dai and Blandford, 2013, Linial and Sari, 2017). In particular these and other researchers investigated the driving effect of the strong gravitational field on accretion processes from the object onto a disk or another physical environment around the BH.

Under the assumption that such is the case for the Sgr A* system we calculated in a post Newtonian approximation the radius R of the orbit of the revolving object. In L1, however, these calculations did not take into account the GR apsidal precession of the orbit. Here we derive again the value of the mean radius of the orbit, assuming that the observed period P1 is the time measured in the observer frame of reference between 2 successive passages of the object in the pericentre point of the orbit.

For an orbit of a small mass $m_P$ relative to the mass $M_{BH}$ of the central BH that is slightly eccentric, there are 3 parameters associated with the revolving object that depend on the mean radius R of the orbit. For the given $M_{BH}$ of the central body and neglecting the mass $m_P$ of the object relative to $M_{BH}$, they are:

1. The Keplerian period $P_k$ in the local frame of reference which is:

(4) $$P_k \cong 2\pi \left[\frac{R^3}{GM_{BH}}\right]^{\frac{1}{2}}$$

2. The GR apsidal precession angle per one revolution in the local frame of reference, which according to Weinberg (1972) is:

(5) $$\delta\varphi = \frac{6\pi G M_{BH}}{c^2 R}$$

In the local frame, the time between successive passages of the pericentre point is

(6) $$P_t = \left(1 + \frac{\delta\varphi}{2\pi}\right) P_k = \left(1 + \frac{3}{2}\frac{R_S}{R}\right) P_k$$

Here $R_s$ is the Schwarzchild radius of the BH:

$$R_S = \frac{2GM_{BH}}{c^2}$$

3. The ratio between the measured period P by a far away observer, and the period in the local frame $P_t$, due to GR+SR time dilation. This is given by:

(7) $$\frac{P_t}{P} = \sqrt{1 - \frac{3R_s}{2R}}$$

Defining $x = \frac{R}{R_S}$ and taking $M_{BH} = 4.28 \times 10^6 M(Sun)$ (Gillessen et al 2016) and $P = 0.1032\ d = 8916\ sec$, from expressions 4, 6 and 7 we get the equation

$$\frac{x^3 \left(1 + \frac{3}{2x}\right)^2}{1 - \frac{3}{2x}} = 563$$

which is satisfied by

$$x = 6.6$$

Thus the proposed object revolves around the SMBH at a distance of 6.6 Schwarzschild radii of the blackhole. Its velocity is about $\frac{1}{3.6}$ of the velocity of light.

The result x=6.6 may perhaps be regarded as an additional, though indirect and not quantifiable evidence for the reality of the periodicity that we find associated with the observed data. Figure 7a1 shows that except for the P1 minimum point there is no other outstanding one in the entire range [2-50] $day^{-1}$. As already stated in Section 4, no other outstanding minimum appears also in the frequency range [50-100] $day^{-1}$. The grid of the frequency F1=1/P1=9.6898 $day^{-1}$ is found with its high statistical significance within the small section of the [2-100] frequency interval of f < 20. For a non rotating BH the radius of the last stable circular orbit is $R_{LSCO}$ = $3R_S$, and with $M_{BH}$ = 4.28x10$^6$ M(Sun) the corresponding Keplerian frequency is $\cong$ 20 $day^{-1}$. Thus more than the 80% of the search interval that include all f >20 values, contain frequencies that have no immediate physical meaning within the context of the BH of Sgr A*. Yet there is no mathematical reason why the best grouping of the 71 numbers could not be found around tick marks of grids of frequencies within this much larger frequency range. The fact that the period that was found is in the physical meaningful section of the search interval, with the a-priori probability to be found there that is less than 0.2, is lending some additional qualitative credibility to the reality of P1, and possibly to our interpretation of it as well.

### 7.3 The nature of the object

If the origin of the pacemaker in the Sgr A* system is indeed a motion of a certain mass in a nearly Keplerian orbit around the BH, it might be a low mass star. The feasibility of this hypothesis has recently received some supporting evidence from the growing observational evidence for cusp in the stellar population around the Milky Way's central black hole (e.g. Sch\"odel, Amaro-Seoane, Gallego-Cano, 2017). More specifically, observations presented recently testify on an ongoing low-mass star formation near Sgr A* (Yusef-Zadeh, F., et al , 2017). The objects that these authors have identified lie 2 orders of magnitude further away from the BH than the few gravitational radii of our object, but it seems that it is the angular resolution limits of the observations that set the scale of distances of these newly discovered low mass stars.

Mass transfer ensues when a stellar (or planetary) component is driven sufficiently close to the black hole, and its Roche lobe is filled. As in L1 we use Eggleton's (1983) expression (2) for the radius $R_L$ of the equi-potential Roche Lobe surface of mass m in a circular orbit of radius R around mass $M_{BH}$, as a function of the mass ratio $q = \frac{m}{M_{BH}}$ :

(8) $$R_L = \frac{0.49 q^{\frac{2}{3}}}{0.6 q^{\frac{2}{3}} + \ln(1+q^{\frac{1}{3}})} R$$

With the value $6.6 R_S$ for the radius of the orbit of a star of mass m=0.18M(Sun), we obtain for the radius of the equi-potential Roche lobe of the star $R_L = 0.2 R(Sun)$. According to the mass-radius relation of low mass stars presented in Fig. 8 in Demory et al. (2009) , this is the radius of a star of mass $0.18 M(Sun)$. We may therefore hypothesize the presence of a star of this mass that nearly fills its Roche lobe surface in a slightly eccentric Keplerian orbit around the SMBH. In other words, the orbit of the star is just at the Tidal Disruption radius of the BH for such a star (Rees, 1988; Mainetti, et al, 2017).

## 7.4 Second Proposition

Here we propose another possible candidate for the revolving object around the BH. Its mass $m_P$ is in the scale of a few percent of Earth mass $m_E$, rather than in the scale of stellar masses: $m_P = \varepsilon m_E = \varepsilon \times 6 \times 10^{27} gr$. With $M_{BH}$ as above we have $q = 7 \times 10^{-13} \varepsilon$.

For the Roche lobe radius $r_L$, expressed in units of Earth radius, its dependence on $\varepsilon$ is given by expression (8) as:

$$r_L = \frac{5.1 \times 10^{-5} \varepsilon^{\frac{2}{3}}}{4.73 \times 10^{-9} \times \varepsilon^{\frac{2}{3}} + \ln(1 + 8.88 \times 10^{-5} \times \varepsilon^{\frac{1}{3}})}$$

On the basis of measurements in 274 exoplanets, Bashi et al (2017) have recently found for low mass planets the Mass-Radius relation

$$R_P \cong m_P^{0.55}.$$

Expressed in Earth units this reads as:

$$r_P \cong \varepsilon^{0.55}$$

Figure 8 is a plot of the two radii $r_L$ and $r_P$ as functions of $\varepsilon$. The two curves intersect around the values $\varepsilon = 0.077$. This means that the radius of the Roche lobe of a planetary object of mass around 8% of Earth mass, that revolves around the SPBH of Sgr A* at a distance of $6.6 R_s$, is equal to the radius of the planet itself. Such an object could be part of the debris of a tidally disrupted star by the intense gravitational field of the BH at some earlier historic times (Frank & Rees, 1976; Coughlin, Nixon, Begelman, and Armitage, 2016).

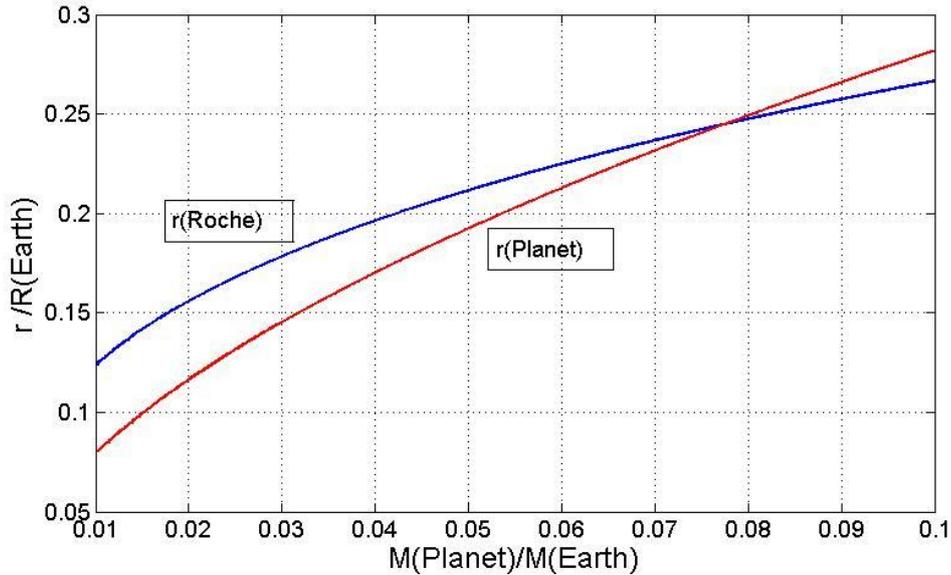

Figure 8: The radius of a planet and the radius of the Roche lobe of a planetary mass at a distance of 6.6 $R_S$ from the BH of Sgr A*, relative to Earth radius, as functions of the mass of the planet relative to Earth mass.

In a slightly eccentric orbit, the radius of the Roche lobe assumes a minimum value when the object passes through the pericentre point of the orbit. In the binary system of the SMBH+planet or star, each passage through the pericentre point excites and enhances tidal instabilities in the outer layers of the object. As already mentioned above, according to Markoff (2010) , Li et al (2015) or N13, the mean frequency in the occurrence of a large X-ray flares of Sgr A* is about 1 or 1.1 flares per day. This would then mean that after some 9 such passages on the average, the tidal waves at the surface of the object are strong enough to cause a large overflow of material through the $L_1$ point of the system onto the central BH. These intense mass loss episodes provide the energy source for the observed large flares of the system. The synodic periodicity of the object in the precessing nearly Keplerian orbit serves as the pacemaker that tightly groups the flares around the tick marks of the 0.1032 d grid on the time axis .

## 8. Summary

The timings of 71 large X-ray flares of the SMBH Sgr A*, recorded by $Chandra$ and $XMM-Newton$ in the 15 years between 2000 and 2014, are modulated by a pacemaker of the period P1=0.103203 day. This result is obtained from an analysis of the time series of the midpoints of the flares with a statistical significance of no less than $3.27\sigma$, and very likely at a $4.62\sigma$ significance level. If it is not a random statistical fluke, or an artifact related to the measurement process, it may be understood as a signal from an object revolving around the BH in a slightly eccentric nearly Keplerian orbit. The radius of the orbit is about 6.6 Schwarzschild radii of the BH. The object might be a low mass star of 0.18 M(Sun) or a small planet of mass that is about 8% of Earth mass.

The reality of a long term operation of a pacemaker with a period P1=0.1032 day in the Sgr A* system should be further tested by future observations in X-rays, as well as in the NIR radiation of the object.

### Acknowledgment

I thank the referee, Dr. Phil Uttley, for making thoughtful comments and suggestions that helped improving this paper considerably.

### References

Abramowicz, M.A. , Fragile, P.C. , 2013. Liv. Rev. Relat. 16, 1
Bashi D., Helled R., Zucker S., and Mordasini C., 2017, A&A, arXiv:1701.07654
Coughlin E.R., Nixon C., Begelman M.C. and Armitage P.J., 2016, MNRAS, 459, 3089
Demory, B.O. , Segransan, D. , Forveille, T. , et al. , 2009. A&A 505, 205
Eggleton, P.P. , 1983., ApJ 26 8, 368
Frank j. & Rees M.J., 1976, MNRAS, 176, 633
Gillessen S., Palewa P. Eisenhauer F. et al, 2016, arXive 1611.09144
Li, Ya-Ping, Yuan, Feng, Yuan Qiang, et al, 2015, ApJ,810, 19L
Li Y.P, Yuan F., Wang D., 2017, MNRAS, 468, 2552
Linial, I., Sari, R., 2017, MNRAS, arXive:1705.01435
Leibowitz, E., 2017, New Astronomy, 50, 52 (L1)

Mainetti D., Lupi A., Campana S., Colpi M., Coughlin E.R., Guilcochon J. and Ramirez-Ruiz E., 2017, A&A, arXiv:1702.07730
Markoff, S. 2010, PNAS, 107, 7196
Meyer, L., Do, A., Ghez, M.R., Morris, S., et al, 2009, ApJ, 694, L87
Mossoux E., Grosso N., 2017, arXive:1704.08102 [astro-ph.HE] 26 April 2017
Neilsen, j. , Nowak, M,A , Gammie, C. , et al. , 2013. ApJ 774, 42
Ponti, G. , De Marco, B. , Morris, M.R. , et al. , 2015, MNRAS 454, 1525
Ponti, G., George, E., Scaringi, S., Zhang,S., Jin,C., et al, 2017, MNRAS, 468, 2447
Rees M., 1988, Nature, 333, 523
Sch\"odel R., Amaro-Seoane P., Gallego-Cano E., 2017, arXiv:1702.00219
Weinberg S., 1972, "Gravitation and Cosmology", John Wiley & Sons, Inc. p. 194
Yusef-Zadeh, F., Cotton W., Wardle M., Royster M.J., Kunneriath D., Roberts D.A.,
    Wootten A. and Sch\"odel R., 2017, MNRAS, arXiv:1701.05939
Zhang, S., Baganoff, F.K., Ponti, G., et al, 2017, arXive:1705.08002 [astro-ph.HE] 22 May 2017